# *In vitro* Demonstration of Cancer Inhibiting Properties from Stratified Self-Organized Micro-Discharge Plasma-Liquid Interface

Zhitong Chen[a], Shiqiang Zhang[a], Igor Levchenko[b], Isak I. Beilis[c], Michael Keidar[a,]*

[a] Department of Mechanical and Aerospace Engineering, The George Washington University, Washington, DC 20052, USA

[b] School of Chemistry Physics and Mechanical Engineering, Queensland University of Technology, Brisbane QLD 4000, Australia

[c] School of Electrical Engineering, Tel Aviv University, Ramat Aviv 69978, ISRAEL

Experiments have revealed a nontrivial cancer-inhibiting capability of liquid media treated by the plasma jet capable of forming thinly stratified self-organized patterns at a plasma-liquid interface. A pronounced cancer depressing activity towards at least two kinds of human cancer cells, namely breast cancer MDA-MB-231 and human glioblastoma U87 cancer lines, was demonstrated. After a short treatment at the thinly stratified self-organized plasma-liquid interface pattern, the cancer inhibiting media demonstrate well pronounced depression and apoptosis activities towards tumor cells, not achievable without interfacial stratification of plasma jet to thin (of several μm) current filaments, which therefore play a pivotal (yet still not completely clear) role in building up the cancer inhibition properties. Moreover, thinly stratified, self-organized interfacial discharge is capable to efficiently control the ROS and RNS concentrations in the cancer-inhibiting media, and in particular, abnormal ROS/RNS ratios not achievable in discharges which do not form stratified thin-filament patterns could be obtained. These results were explained in terms of interaction of thin plasma filaments of the self-organized pattern with gas and liquid, where the unusual interaction conditions (i.e., high surface-to-volume ratios etc.) cause accumulation of ROS, RNS and other species in unusual ratios and concentrations, thus forming potentially efficient anti-cancer cocktail. Our funding could be extremely important for handling the cancer proliferation problem, and hence, it should be brought to light to attract due attention of the researchers and explore the possible potential of this approach in tackling the challenging problem of high cancer-induced mortality and rising morbidity trends.

---

* Corresponding Author:
E–mail address: zhitongchen@gwu.edu (Z. Chen), gor.Levchenko@qut.edu.au (I. Levchenko), beilis@eng.tau.ac.il (I. Beilis), keidar@gwu.edu (M. Keidar)

# 1. Introduction

Despite tremendous efforts undertaken, cancer and cancer-related diseases still remain among the most dangerous and mortiferous abnormalities responsible for about 13% of human death cases, totally accounting for more than 7 million per year [1]. Moreover, cancer morbidity tends to rise and about 11 million deaths are expected in 2030. Undoubtedly, cancer represents a problem of paramount importance, and consolidated, sophisticatedly designed approaches are vitally required to tackle this challenging problem.

Many quite different treatment methods including surgical techniques, medication drugs, and radiation-based approaches are now in an active list, but much advanced progress is needed to cope with this stubborn problem and drastically drop down the mortality rate. Indeed, the problem still persists despite huge effort applied. *Are there any ways to reverse the trend and find a really radical breakthrough? Apparently, novel approaches with unexpected, counterintuitive mechanisms should to be explored.*

Nanoparticle and plasma-based cancer therapy is one of the novel techniques which have recently demonstrated a significant potential in curing various types of cancers [2, 3]. It was demonstrated that some nanomaterials, as well as cold and non-thermal atmospheric-pressure plasma indeed inhibits proliferation of human cancer cells [4, 5], and various mechanisms were proposed to explain this effect including the action of reactive oxygen and nitrogen species (ROS and RNS) [6], DNA damage [7] and others. In addition, it was shown that plasma action can be selective towards the cancer cells while sparing normal cells, which is the main goal of anti-cancer therapies ("Holy Grail").

To move further, *synergetic effects* were examined by involving new generation of nanomaterials such as hierarchically designed nonporous and core-shell nanoparticles [8, 9] and encouraging results were obtained. Another idea that was put forward recently is based on a plasma-stimulated media [10-13] to produce highly active complex material enriched with ROS, RNS and other active species to directly treat the tumor cells. Although efficient in general, this approach still requires further enhancement and specifically, much higher levels of controlling the active species densities and ratios in the plasma-activated therapeutic media. Specific strategies amendable for adaptation of plasma-stimulated media were proposed but not yet realized [14], and among them, the approaches based on miniaturization of discharge (and ultimately, involvement of micro-discharge and micro plasmas), and self-organizational phenomena attract a close attention.

Self-organization is one of the key features of living matter. It is based on an entropy minimum principle and serves as a real engine for the development of animated nature. Moreover, it is also intrinsic to complex non-living systems; albeit not in such ubiquitous amount, self-organization still plays important role in some physical systems. An atmospheric pressure glow micro-discharge plasma (μAPGDP) with self-organized thinly stratified interface patterns is a characteristic example of such a self-organizing system [15, 16] which was utilized in this work to prepare the cancer-killing plasma-activated therapeutic media.

In this work, we present a novel approach which demonstrates nontrivial cancer-inhibiting capabilities of *spontaneous pattern-forming self-organization at the interface* between atmospheric pressure glow discharge plasma and liquid media. Specifically, in certain discharge modes (i.e., current-voltage and pressure conditions), the solid plasma jet encountering liquid surface forms a self-organized interfacial pattern where the bulk plasma is stratified to thin (up to μm in diameter at the tops, see **Figure 1** for the general schematics and the relevant photographs) current filaments which imprint complex shapes on the surface (See also a movie available at the journal website). After a short treatment at the self-organized plasma-liquid interface pattern, the cancer-inhibiting media has acquired a pronounced cancer-depression activity towards at least two kinds of human cancer cells, namely breast cancer MDA-MB-231 and human glioblastoma U87 cancer lines. Most important, the cancer-inhibiting media demonstrate tumor cell depression and apoptosis activities not achievable without interfacial self-organization and effect of bulk plasma jets not containing thin, μm-scaled current filaments. Apparently, the complex stratified self-organized interfacial patterns play a pivotal role in building up the cancer-inhibition properties, but the specific mechanism and relation between inhibition of the cancer proliferation and apoptosis is still not completely clear and more efforts should be applied to fully discover the possible potential of this approach in tackling the challenging problem of high cancer-induced mortality and rising morbidity trends. A plausible mechanism is also discussed in terms of interaction of thin plasma filaments with gas and liquid causing accumulation of ROS, RNS and other species in unusual ratios and concentrations, forming potentially efficient anti-cancer cocktail.

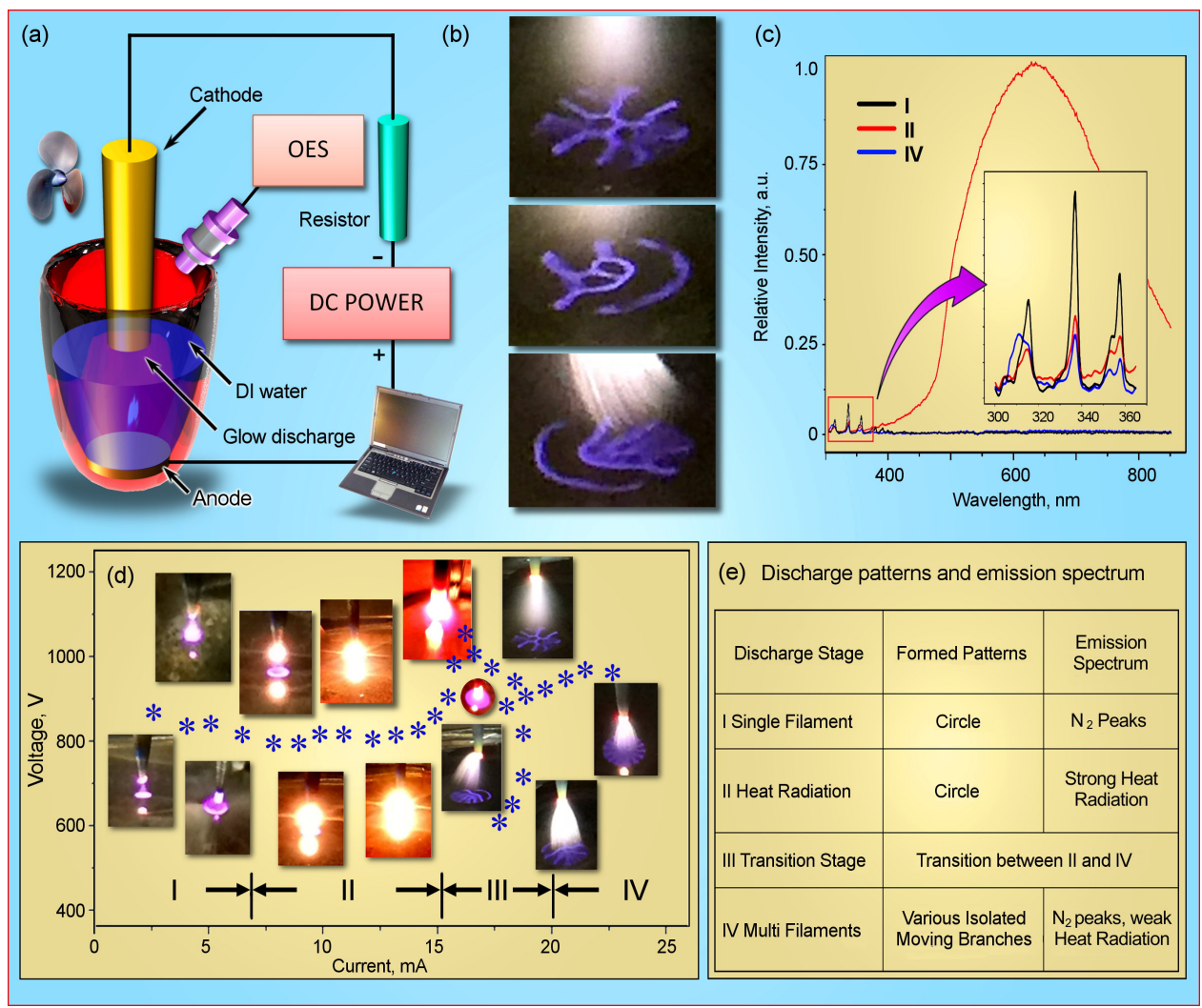

**Figure 1**. (a) Schematic representation of the atmospheric pressure glow micro-discharge setup. A small (several mm) gap between the cathode and surface of liquid accommodated a bunch of plasma. (b) Optical photographs of the discharge patterns above the therapeutic media during the activation process. The self-organized patterns have complex structure strongly depending on the voltage-current conditions. Stratification of plasma jet to a large number of thin current filaments is clearly visible at lower photograph of the panel; fine complex patters are also formed on the surface of liquid media; (c) Optical emission spectrum by the atmospheric pressure glow discharge above DI water taken using UV-visible-NIR in the 200-850 nm wavelength range. (d) Current-voltage dependence of the system with optical photographs of the self-organized stratified interface patterns. Four discharge stages could be clearly determined. (e) A table summarizing the relations between discharge stages, self-organized patterns and optical emission.

## 2. Results and discussion

### 2.1. Self-organization and plasma

Atmospheric pressure glow discharge and micro-discharge plasma with self-organized patterns is a growing research area attracting strong attention of experts in various fields to deeply explore the physical mechanisms behind the self-organization [17, 18]. Apart from the fascinating phenomena, the effect of μAPGDP on tumor and plasma-activated cancer therapy media are among the most important features of this system. Specifically, we have used here the atmospheric pressure micro-discharge plasma with self-organized stratified interface patterns on a liquid surface (deionized water, DI) to produce the plasma-activated therapeutic media under various regimes, and apply them to human breast cancer cells line (MDA-MB-231) and glioblastoma cancer line (U87) to conceptually reveal potential of this novel media/material in cancer therapy, and in particular, to demonstrate that the stratified interface pattern on a liquid surface plays a key role in the activation of therapeutic media.

Discharge mode transitions are shown to be leading to alteration of plasma-stimulated media to affect cancer cells. Moreover, plasma-activated therapeutic media has demonstrated more pronounced killing effect on glioblastoma cancer cells than breast cancer cells. This is largely unexpected results based on previous experience with these cell lines and their response to cold atmospheric plasma.

## 2.2. Self-organized stratified patterns at plasma/liquid interface

The self-organized patterns and other phenomena are widely observed in both natural and anthropic fields, within diverse types of chemical, biological and physical systems including low temperature plasma [19, 20]. Atmospheric pressure glow micro-discharge plasma with self-organized stratified interface pattern has gained an increasing attraction due to its unique features, ease ignition and maintenance, and diverse emerging applications [17]. Various plasma devices are able to produce self-organized plasma patterns, including DC micro-discharge devices with gas feed [21], a pin water-anode glow discharge setups [22], and dielectric barrier discharge-like devices [23], Many key parameters such as gap length, excitation frequency, applied voltage etc. were examined to study the effect on the pattern formation [24, 25]. Moreover, numerical simulations were utilized to explain the basic physical mechanisms behind the formation of self-organized patterns [17, 26].

We stress here that the μAPGDP with the self-organized stratified interface patterns for the cancer cell treatment applications has not been reported yet, although atmospheric pressure plasma studies for the biomedical applications and cancer therapy by reactive oxygen and nitrogen species (ROS and RNS) [27] has been reported and demonstrated some potential.

## 2.3. Results of self-organized plasma experiment

Fig. 1b shows optical photographs of various self-organized stratified interface patterns with micro-discharges. Complex shapes consisting of radial and confocal lines of different density may be produced. Some elements of symmetry (both radial and axial) can be sometimes observed.

The current-voltage characteristics and optical photographs discharge patterns above the DI water corresponding to the specific current/voltage conditions are shown in Fig. 1d. The whole current-voltage characteristic can be divided to the four specific stages. At stage I (current not exceeding 7 mA) the discharge voltage and current are low, and the discharge pattern represent a single filament (i.e., contracted glow discharge) following the initial corona discharge. As the discharge voltages increases to certain degree (stage II), the temperature of the tungsten cathode increases, resulting into drastically enhanced heat radiation, as clearly seen in Figure 1c, stage II. At stage III, the discharge enters an unstable state where it alternates between two (II and IV) stages featuring strong heat radiation (stage II) and the multi-filament pattern (stage IV). A plausible reason for this behavior could be that the high temperature results in stronger thermionic rather than the secondary electron emission from a comparatively cold electrode; thus, the discharge flips from the multi-filament and heat radiation-supported stage where the thermionic emission is efficient enough to maintain the increased discharge current. At the IV stage, the discharge stabilizes at the multi-filament stage and stretches out to a number of discharge filaments from the cathode to the liquid media surface. Various types of self-organized stratified interface patterns are formed above the liquid media surface, as shown in Fig. 1b and 1c.

Fig. 1c shows spectrum of the μAPGDP taken from the self-organized stratified interface patterns at the liquid media surface/plasma interface. Measurements were made for the three different stages, namely I, II, and IV (transient stage III is not characteristic due to an unstable current regime). The emission bands were identified by the technique described elsewhere [28]. Species at wavelengths 337, 358, and 381 nm were identified as a low-density N2 second-positive system ($C3\Pi u - B3\Pi g$). The intensity of emission at II stage is much higher than that of I and IV stages (see graphs at Fig. 1d) due to high discharge intensity at II stage.

*Based on the above described considerations, we have selected stages (modes) I and IV as the basic platforms for the bio-oriented studies described below in detail.*

**2.4 Liquid media activated by self-patterned plasma**

We have studied the effect of atmospheric pressure glow micro-discharge producing self-organized stratified interface patterns on the composition of liquid media at the two major discharge regimes, namely regime I (low-current mode when the discharge current does not exceed 7 mA), and mode IV which is characterized by well-shaped self-organized patterns at the plasma-liquid interface, as shown in Fig. 1c. Apparently, mode II (very bright radiation shown in Fig. 1c) might have very strong heat effect to the liquid medium being activated, leading to a very large number of hardly predictable chemical reactions and eventually, unpredictable solution composition. On the other hand, mode III is unstable and does not represent a specific interest. On the contrast, mode IV which featured very high current but potentially 'soft' effect due to low thermal emission and more importantly, due to the well pronounced self-organized patterns at the interfaces, represent a special interest to the media activation and further, to the effect on tumor cells.

To avoid overheating, relative short treatment times of 12, 24, 36, 48, 60 seconds were chosen (further measurements have confirmed the adequacy of such a choice). Since these measurements are essentially concept-proof studies, the concentrations of the two main reactive species, $H_2O_2$ and $NO_2^-$, mostly important in bio- and medical applications (including cancer-killing therapy) were measured.

We stress there that our study presents the first results of the effect of self-organized plasma with stratified interface patterns on RONS and human cancer cells; thus, we have thoroughly compared the concentrations of $H_2O_2$ and $NO_2^-$ obtained in our experiments with the results described in literature for the different kinds of the discharge, to clearly stress the unique features of the self-organized plasma. The data are also listed in Table 1.

**Table 1**

$H_2O_2$ and $NO_2^-$ concentrations, as well as discharge conditions obtained in these experiments and compared similar results described in publications (experimental and simulations)

| Ref. | $H_2O_2$, mmol/L | $NO_2^-$, mmol/L | Current/energy | Discharge type, working gas |
|---|---|---|---|---|
| This work | 0.015 | 0.12 – 0.7 | 3…20 mA | DC with spatial pattern |
| [29] | 0.7 | 0.2 | 10-15 mA | |
| | 2..8 | 0.23 | 300-400 kJ/L | Pulsed |
| [30] | 0.3 | 0.04 | 100-200 kJ/L | Corona |
| | 0.1 | 0.04 | 300-400 kJ/L | Plasma jet |
| [31] | 9.8 – 1.2 | 0.2 – 0.5 | 10 – 18 A | DC transient spark discharge generated in atmospheric pressure |
| [32] | 0.2 | 0.1 | 20 A | Air |
| [33] | 8 | 2×10-5 | | Simulations |

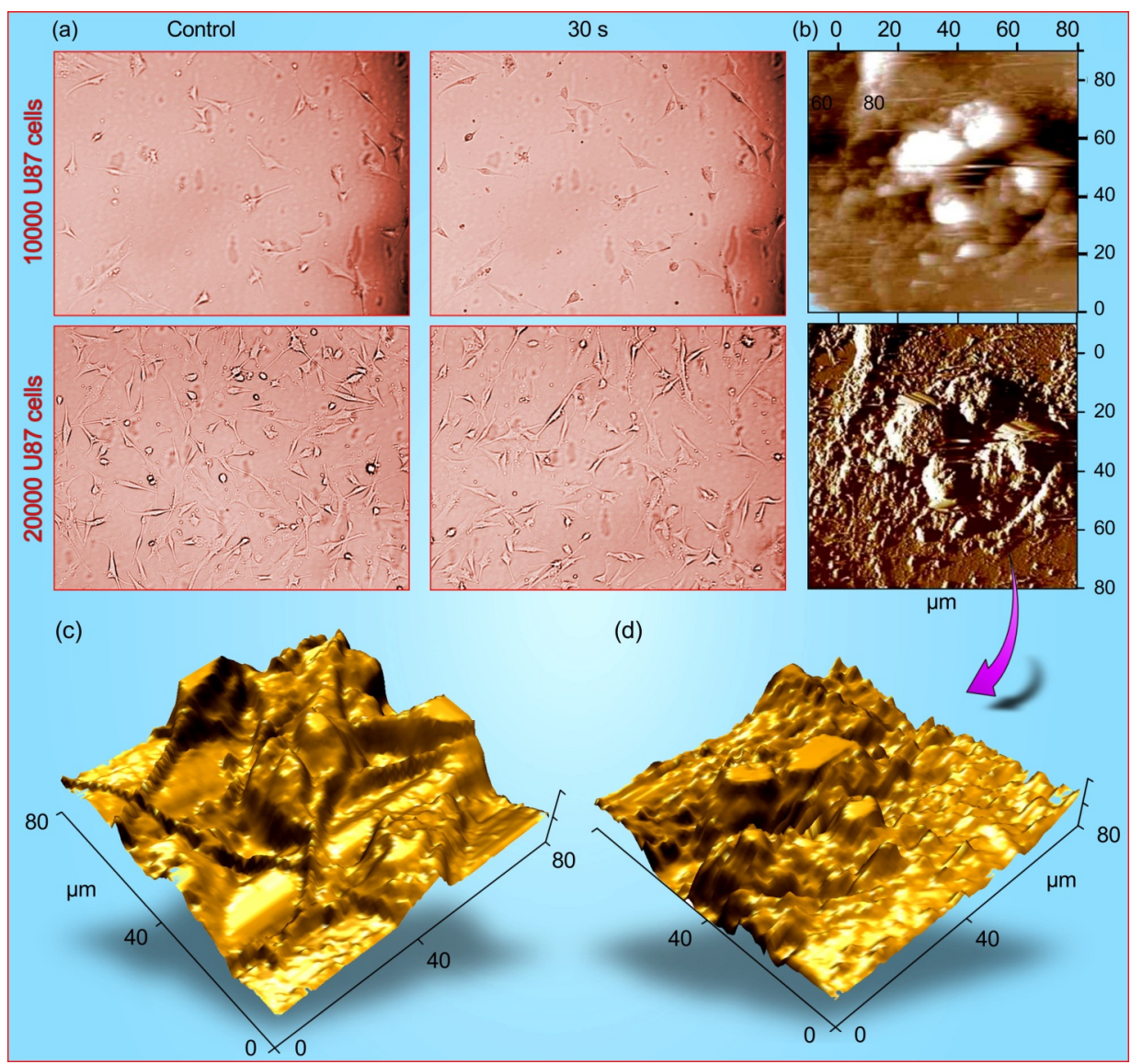

**Figure 2**. (a) Optical microscopy of 10,000 (upper row) and 20,000 (lower row) U87 cells, control and after 30 second treatment, after 24 h of incubation. (b) AFM images of U87 cells after 24 h of incubation and 30 s treatment. (c, d) Three-dimensional reconstruction of U87 cells before treatment (c) and after treatment (d), respectively.

Fig. 2 shows the typical optical and atomic force microscopy characterization results for the untreated and plasma-treated U87 cells. Decrease in number of cells with time, as well as destruction can be observed in these two types of characterization. The three-dimensional reconstructions of U87 cells (Fig. 3c and 3d) show the details of surface morphology of untreated and threated cells.

The ROS and RNS contents in the APGDP-treated liquid media (we recall here that the deionized water was used as a start solution) were determined. The time dependencies of the $H_2O_2$ and $NO_2^-$ concentrations in the APGDP-stimulated liquid media produced by the two (low- and high-current) modes are shown in Fig. 3a and 3b. For the high-current mode (panel A), the concentration of $H_2O_2$ decreases and the concentration of $NO_2^-$ increases with the treatment time. The low-current mode (panel B) demonstrates opposite trend: both $H_2O_2$ and $NO_2^-$ concentrations rise with the time. Absolute concentrations also demonstrate different trends: while hydrogen peroxide $H_2O_2$ concentration features highest (17 μM versus 12 μM) numbers at low-current mode, the $NO_2^-$ concentration reaches strong maximum (650 μM versus 120 μM) under the high-current mode.

Examination of the processes taking place at the plasma-liquid interface allows suggesting the following mechanisms of $H_2O_2$, $NO_2^-$ and other RONS formation in plasma-stimulated media [34-36]:

$$O_2 + e^- \rightarrow O_2^- \tag{1}$$

$$O_2^- + H^+ \rightarrow HO_2 \tag{2}$$

$$HO_2 + e^- \rightarrow HO_2^- \tag{3}$$

$$HO_2^- + H^+ \rightarrow H_2O_2 \tag{4}$$

$$H_2O + e \rightarrow H_2O^* + e \tag{5}$$

$$H_2O^* \rightarrow {}^*OH + H^* \tag{6}$$

$$H^* + O_2 \rightarrow {}^*OH + O \tag{7}$$

$$O_2 + e \rightarrow O + O + e \tag{8}$$

$$O + H_2O \rightarrow 2\,{}^*OH \tag{9}$$

$$^*OH + O \rightarrow {}^*HO2 \tag{10}$$

$$^*HO_2 \rightarrow H+ + O_2^{*-} \tag{11}$$

$$O_2^{*-} + O_2^{*-} + 2H_2O \rightarrow H_2O_2 + O_2 + 2OH^- \tag{12}$$

$$O_2^{*-} + O_2^{*-} + 2H^+ \rightarrow H_2O_2 + O_2 \tag{13}$$

$$O + O_2 \rightarrow O_3 \tag{14}$$

$$^*OH + {}^*OH \rightarrow H_2O_2 \tag{15}$$

Other possible reactions illustrating the routes of $H_2O_2$ and $NO_2^-$ formation in liquid and plasma are listed elsewhere [37]. According to these mechanisms, the concentration of $H_2O_2$ and $NO_2^-$ should increase with treatment time, while the concentration of $H_2O_2$ (at high current) was actually decreasing. The possible reason for this behavior is that the temperature of plasma at the high current mode is much higher than that at the low current conditions; since hydrogen peroxide is thermodynamically unstable, its rate of decomposition increases with rising temperature [38].

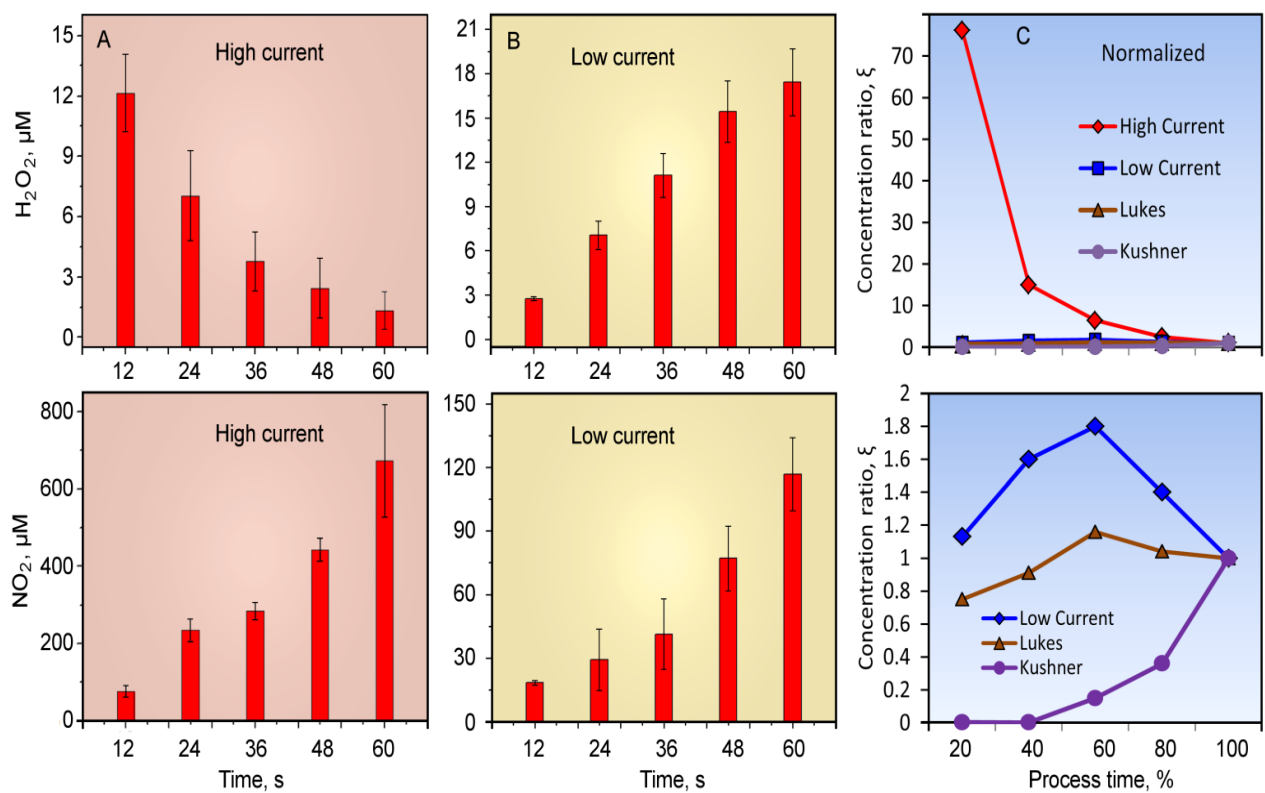

**Figure 3**. Time dependencies of ROS and RNS concentrations in liquid media treated with self-organized micro-discharge forming stratified interface patterns. (a) $H_2O_2$ concentration and $NO_2^-$ of high current (IV stage); (b) $H_2O_2$ concentration and $NO_2^-$ of low current (I stage). The concentration of $H_2O_2$ decreases and the concentration of $NO_2^-$ increases with the treatment time in high-current mode, while both $H_2O_2$ and $NO_2^-$ concentrations rise with the time. The absolute concentration of $H_2O_2$ features highest numbers at low-current mode, while the $NO_2^-$ concentration reaches strong maximum under the high-current mode. (c) Normalized (upper panel) and absolute (lower panel) $H_2O_2$ / $NO_2^-$ ratios for the result obtained in this work, as well as in Refs. 31 and 32, for comparison.

Fig. 3c shows the most unusual features of the ROS and RON measurements in liquid media stimulated by the self-organized plasma with stratified interface patterns. Whereas the normalized (upper panel) $H_2O_2/NO_2^-$ ratios for the result obtained in this work demonstrate strong drop with the process time (i.e., the $H_2O_2$ concentration significantly exceeds that of $NO_2^-$), the results obtained for other kind of plasmas (experimental studies by Machala et. al. [29] and theoretical findings by Takahashi et. al. [30], see also Table 1) demonstrate the numbers between zero and 1. The absolute graphs for the same ratio are shown in the lower panel. One can see that the ratios obtained in our experiments (up to 70) are very unusual in comparison with other kinds of plasma.

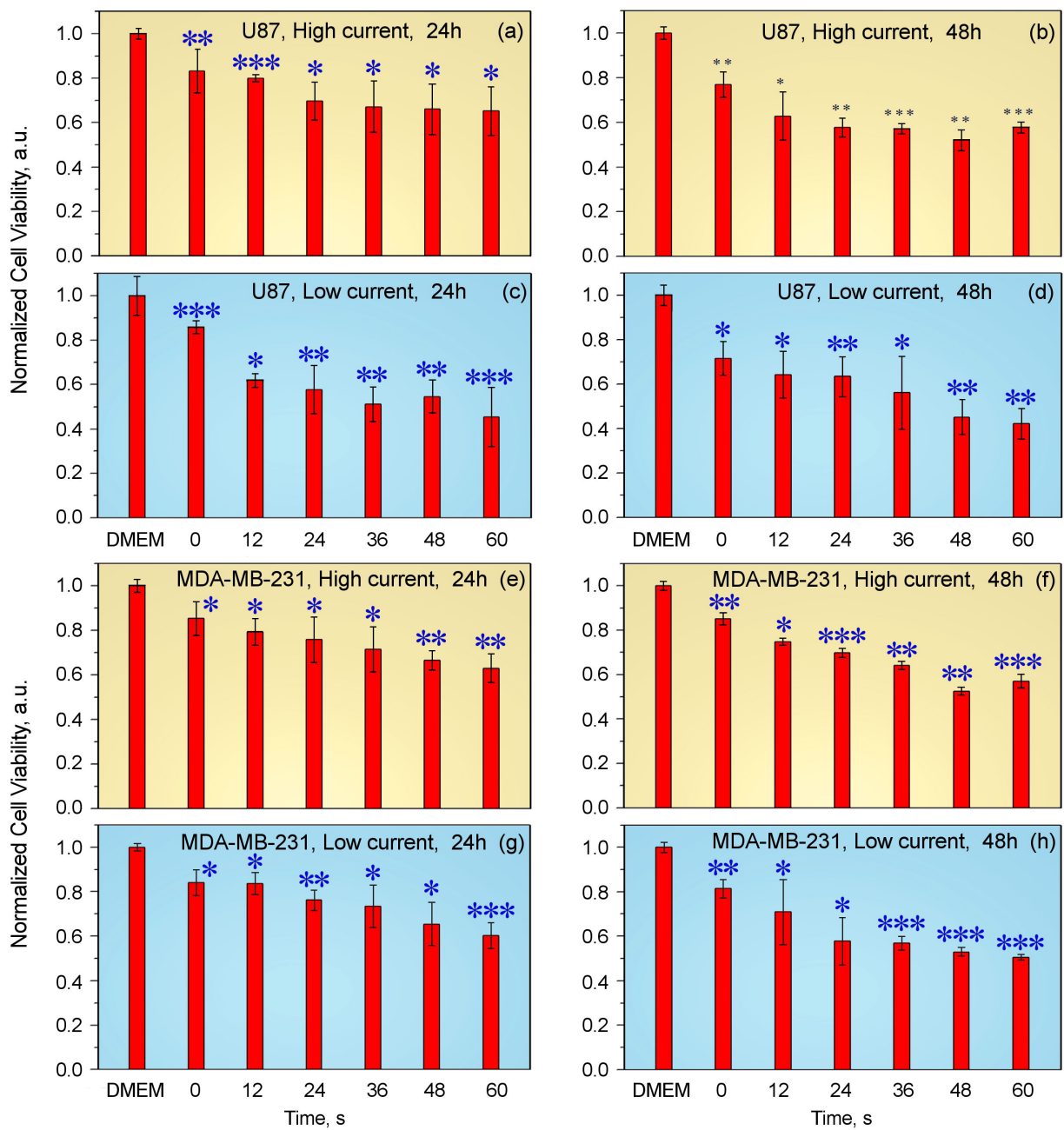

**Figure 4**. Effects of seven media: DMEM, DI water (0 second treatment), and five plasma-produced activated media (DI water activated by APGDP at high and low currents for 12, 24, 36, 48, and 60 seconds on viability of the U87 human glioblastoma cancer cells (a-d) and MDA-MB-231 human breast cancer cells (e-h) after 24 and 48 hours incubation. The percentages of surviving cells for each cell line were calculated relative to controls (DMEM). Student's t-test was performed, and the significance compared to the first bar is indicated as *$p < 0.05$, **$p < 0.01$, ***$p < 0.005$ (n=3).

## 2.5. Effect of plasma-activated media onto human cancer cells

To study the effect of the liquid media stimulated by the plasma with self-organized stratified interface pattern, two kinds of cancer cells were treated. The APGDP-stimulated media was applied to the human glioblastoma (U87) and breast (MDA-MB-231) cancer cells, whereas the Dulbecco's Modified Eagle's Medium (DMEM) was used as the control platform Fig. 4a - 4d show the viability of human glioblastoma cancer (U87) exposed to DMEM, deionized (DI) water (i.e., 0 second treatment), and liquid media treated by self-organized plasma at high (Fig. 4a and 4b) and low (Fig. 4c and 4d) currents for 12, 24, 36, 48, and 60 seconds, incubated for 24 h (Fig. 4a and 4b) and for 48 h (Fig. 4c and 4d). Liquid media (30%) was mixed with DMEM (70%) applied to cells. The viability of the glioblastoma cancer cells incubated for 24 h in media treated in plasma at low current was lower than the viability of cells incubated in media treated at high current. The viability of glioblastoma cancer cells incubated for 48 h in high current treated media decreased by approximately 23.0, 37.2, 42.3, 42.8, 48.0 and 42.4%. For the case of low current treated media, the viability of cells decreased by approximately 28.4, 35.8, 36.5, 43.7, 54.8, and 57.8%, respectively. The minimum viability of U87 cells was detected for the media treated by high current (mode IV - self-organization at the liquid-plasma interface) incubated for 48 seconds. In all experiments with the U87 cells the viability decreased with the increase of the plasma treatment time.

Fig. 4e - 4h show the viability of the human breast cancer cells (MDA-MB-231) exposed to DMEM, deionized water, and plasma-activated media after incubation for 24 and 48 h. The viability of breast cancer cells incubated in plasma media activated at low current (i.e., in mode I without self-organize patterns) is in general lower than that of cells incubated in media activated by plasma at high current (in the self-organized pattern mode). The viability of breast cancer cells incubated for 48 h in media activated at low current always decreases with the treatment duration, while the viability of cells incubated in media activated at high current firstly decreases, and then increases a bit.

The comparison of the effect of plasma-stimulated media on human breast and glioblastoma cancer cells reveals more pronounced effect on the viability of glioblastoma than breast cancer cells. It's known that ROS can induce apoptosis and necrosis, whereas RNS induces damage to DNA resulting into cell death [39, 40]. Hence, the trend of both U87 and MDA-MB-231 cancer cells death after incubation on the media activated by self-organized plasma at low current can be attributed to the increase of ROS and RNS concentrations with the treatment time. Moreover, a synergistic effect of RNS and ROS could play an important role in the apoptosis effect [41]. When the therapeutic media is processed at high current (i.e., self-organized plasma patterns are established at the plasma-liquid interface), the viability trend for both types of cells might depend on ROS more than RNS, because the highest concentration of RNS did not result into lowest cell viability. Comparing the viability behavior for both cell types incubated in the high- and low current - activated media, we can indicate that the elevated ROS concentration may play more important role than the RNS-induced apoptosis. Besides, we should point out that lower viability of U87 cells versus that of breast cancer cells is an unexpected and very important funding (given that the U87 cells are considered to be more resilient), and thus, more detailed studies should be encouraged.

## 3. Conclusion

In summary, this paper presents the results of new studies revealing the role and significant potential of self-organization at the interface between atmospheric microplasma and liquid plasma-activated media capable of efficiently inhibiting the growth and proliferation of at least two kinds of human cancer cells, namely breast cancer MDA-MB-231 and human glioblastoma U87 cancer lines. Based on the current-voltage characteristics of the micro-discharge and optical self-organized patterns at the liquid surface, we have defined the four quite different micro-discharge modes and have demonstrated that the activation under self-organized conditions plays a pivotal (yet still not completely clear) role in the synthesis of novel cancer-active media. Thus, we have discovered a novel, unexplored mechanism featuring pronounced capabilities for tumor inhibition, which could have a great anti-cancer potential but still is not well understood. Moreover, we have demonstrated that the self-organized micro-discharge is capable of efficient controlling the ROS and RNS concentrations in the therapeutically media, and in particular, the ROS/RNS ratios not achievable by other types of discharges could be obtained. Our funding could be extremely important for handling the cancer proliferation problem, and hence, it should be brought to light to attract due attention of the researchers and explore the possible potential of this approach in tackling the challenging problem of high cancer-induced mortality and rising morbidity trends.

## 4. Experimental Section

*Novel self-patterned glow micro-discharge plasma device:* Fig. 1a shows a schematic representation of the glow atmospheric pressure micro-discharge setup capable of producing well-defined self-organized stratified interface patterns at the liquid media surface/plasma interface. The media-producing discharge and self-organized interfacial patterning were organized as follows. Anode (thin copper plate, thickness d = 0.2 mm, Ø = 22 mm) was placed at the bottom of a glass-made treat well. Above the plate, 6 ml of deionized water was added to the well. The tungsten cathode of Ø = 2 mm was then installed above the water surface. A ballast resistor (90 kΩ) was connected between the cathode and the direct current (DC) power supply unit (Power Design, Model 1570A, 1-3012V, 40 mA). Voltage is applied between cathode and liquid-immersed anode, and a small (1-2 mm) gap between the cathode and surface of liquid accommodated a bunch of plasma. DI water was treated by self-patterned glow discharge plasma with 12, 24, 36, 48, and 60 seconds to obtain plasma solutions applied to cancer cells.

*Optical emission spectra measurement:* UV-visible-NIR, a range of wavelength 200-850 nm, was investigated on plasma to detect various RNS and ROS (nitrogen [$N_2$], nitric oxide [–NO], nitrogen cation [$N^{+2}$], atomic oxygen [O], and hydroxyl radical [–OH]). The spectrometer and the detection probe were purchased from Stellar Net Inc. In order to measure the radius of the plasma in DI water, a transparent glass plate was used to replace part of container. The optical probe was placed at a distance of 2 cm in front of plasma jet nozzle. Integration time of the collecting data was set to 100 ms.

*Cell culture:* The human breast cancer cell line (MDA-MB-231) and glioblastoma cancer cell line (U87) were provided by Dr. Zhang's lab and Dr. Young's lab at the George Washington University, respectively. Both cells were cultured in Dulbecco's Modified Eagle Medium (DMEM, Life Technologies) supplemented with 10% (v/v) fetal bovine serum (Atlantic Biologicals) and 1% (v/v) penicillin and streptomycin (Life Technologies). Cultures were maintained at 37 °C in a humidified incubator containing 5% (v/v) $CO_2$.

*Evaluation of ROS concentration:* Fluorimetric Hydrogen Peroxide Assay Kit (Sigma-Aldrich) was used for measuring the amount of $H_2O_2$. A detailed protocol can be found on the Sigma-Aldrich website. Briefly, we added 50 $\mu$l of standard curves samples, controls, and experimental samples (DI water treated by self-patterned glow discharge plasma with 12, 24, 36, 48, and 60 seconds) to the 96-well flat-bottom black plates, and then added 50 $\mu$l of Master Mix to each of wells. We incubated the plates for 20 min at room temperature protected from light on and measured fluorescence by Synergy H1 Hybrid Multi-Mode Microplate Reader at Ex/Em: 540/590 nm.

*Evaluation of RNS concentration:* RNS level were determined by using the Griess Reagent System (Promega Corporation) according to the instructions provided by the manufacturer. Briefly, we added 50 $\mu$l of standard curves samples, controls, and experimental samples to the 96-well flat-bottom plates. Then dispense 50 $\mu$l of the Sulfanilamide Solution to all samples and incubate 5-10 minutes at room temperature. Finally, dispense 50 $\mu$l of the NED solution to all wells and incubate at room temperature 5-10 minutes. The absorbance was measured at 540 nm by Synergy H1 Hybrid Multi-Mode Microplate Reader.

*Characterization cell viability of MDA-MB-231 and U87:* The cells were plated in 96-well flat-bottom microplates at a density of 3000 cells per well in 100 $\mu$L of complete culture medium. Cells were incubated for 24 hours to ensure proper cell adherence and stability. On day 2, 40 $\mu$L of and experimental samples (DI water treated by self-patterned glow discharge plasma with 12, 24, 36, 48, and 60 seconds) were added to cells, and 40 $\mu$l of DMEM and DI water was added to the cells. Cells were further incubated at 37 °C for 24 and 48 hours. The cell viability of the glioblastoma and breast cancer cells was measured for each incubation time point with an MTT assay. 100 $\mu$L of MTT solution (3-(4, 5-dimethylthiazol-2-yl)-2,5-diphenyltetrazolium bromide) (Sigma-Aldrich) was added to each well followed by 3-hour incubation. The MTT solution was discarded and 100 $\mu$L per well of MTT solvent (0.4% (v/v) HCl in anhydrous isopropanol) was added to the wells. The absorbance of the purple solution was recorded at 570 nm with the Synergy H1 Hybrid Multi-Mode Microplate Reader.

*Statistical analysis:* All results were presented as mean ± standard deviation plotted using Origin 8. Student's t-test was applied to check the statistical significance (*$p<0.05$, **$p<0.01$, ***$p<0.001$).

## Acknowledgements

*This work was supported in part by a National Science Foundation, grant 1465061. We thank Dr. Ka Bian from Department of Biochemistry and Molecular Medicine at The George Washington University for their support in the measurement of ROS and RNS experiments. I. L. acknowledges the support from the School of Chemistry, Physics and Mechanical Engineering, Science and Engineering Faculty, Queensland University of Technology.*